\documentclass[10pt,superscriptaddress,nofootinbib,notitlepage,tightenlines,twocolumn, pra]{revtex4-1}

\usepackage{amsthm,amsmath,amssymb}

\usepackage{mathrsfs}
\usepackage{amsthm}
\usepackage{amsmath}
\usepackage{amssymb}
\usepackage{graphicx}
\usepackage{epstopdf}
\usepackage{url}
\usepackage{float}
\usepackage{bm}
\usepackage{ifthen}
\usepackage[usenames,dvipsnames]{color}
\usepackage{mathrsfs}
\usepackage[colorlinks=true,citecolor=blue,urlcolor=black]{hyperref}
\usepackage{float}

\newcommand{\be}{\begin{equation}}
\newcommand{\ee}{\end{equation}}

\newcommand{\rv}[1]{{\bf{#1}}}

\newcommand{\sket}[1]{{\ensuremath{\lvert#1\rangle}}}
\newcommand{\lket}[1]{{\ensuremath{\left\lvert#1\right\rangle}}}
\newcommand{\ket}[1]{\if@display\lket{#1}\else\sket{#1}\fi}

\newcommand{\sbra}[1]{{\ensuremath{\langle#1\rvert}}}
\newcommand{\lbra}[1]{{\ensuremath{\left\langle#1\right\rvert}}}
\newcommand{\bra}[1]{\if@display\lbra{#1}\else\sbra{#1}\fi}

\newcommand{\sbraket}[2]{{\ensuremath{\langle#1\rvert#2\rangle}}}
\newcommand{\lbraket}[2]{{\ensuremath{\left\langle#1\!\left\rvert\vphantom{#1}#2\right.\!\right\rangle}}}
\newcommand{\braket}[2]{\if@display\lbraket{#1}{#2}\else\sbraket{#1}{#2}\fi}

\newcommand{\sketbra}[2]{{\ensuremath{\lvert #1\rangle\!\langle #2\rvert}}}
\newcommand{\lketbra}[2]{{\ensuremath{\left\lvert #1\right\rangle\!\!\left\langle #2\right\rvert}}}
\newcommand{\ketbra}[2]{\if@display\lketbra{#1}{#2}\else\sketbra{#1}{#2}\fi}

\newcommand{\cX}{\mathcal{X}}

\newcommand{\cZ}{\mathcal{Z}}

\newcommand{\rvS}{\textbf{S}}

\theoremstyle{plain}

\theoremstyle{definition}

\begin{document}

\title{Experimental composable security decoy-state quantum key distribution using time-phase encoding}

\author{Hua-Lei Yin}\email{hlyin@nju.edu.cn}
\affiliation{National Laboratory of Solid State Microstructures and School of Physics, Nanjing University, Nanjing 210093, China}
\affiliation{Zhongchuangwei Quantum Co., Ltd., Beijing 101400, China}
\author{Peng Liu}
\author{Wei-Wei Dai}
\author{Zhao-Hui Ci}
\affiliation{Zhongchuangwei Quantum Co., Ltd., Beijing 101400, China}
\author{Jie Gu}
\affiliation{National Laboratory of Solid State Microstructures and School of Physics, Nanjing University, Nanjing 210093, China}
\author{Tian Gao}
\author{Qiang-Wei Wang}
\author{Zi-Yao Shen}
\affiliation{Zhongchuangwei Quantum Co., Ltd., Beijing 101400, China}

%%%%%%%%%%%%%%%%%%%%%%%%%%%%%%%%%%%%%%%%%

\begin{abstract}
Quantum key distribution (QKD) promises provably secure communications. In order to improve the secret key rate,
combining a biased basis choice with the decoy-state method is proposed. Concomitantly, there is a basis-independent detection efficiency condition, which usually cannot be satisfied in a practical system, such as the time-phase encoding. Fortunately, this flaw has been recently removed theoretically and experimentally using the fact that the expected yields of single-photon states prepared in two bases stay the same for a given measurement basis. However, the security proofs do not fully consider the finite-key effects  for general attacks. In this work, we provide the rigorous finite-key security bounds for four-intensity decoy-state BB84 QKD against coherent attacks in the universally composable framework. Furthermore, we build a time-phase encoding system with 200 MHz clocked to implement this protocol, in which the real-time secret key rate is more than 60 kbps over 50 km single-mode fiber.
\end{abstract}

\maketitle
\section{Introduction}
Encryption is an important foundation for ensuring information security. With the rapid development of quantum computing technology, current public-key encryption systems will be seriously threatened. Quantum key distribution (QKD) allows two remote users to exchange information-theoretic secure key via the quantum laws. Since the first QKD protocol, BB84, was proposed by Bennett and Brassard in 1984~\cite{bennett1984quantum}. After nearly 40 years of development, BB84 QKD has become the most practical protocol in quantum information science~\cite{liao2018satellite}. By exploiting the weak coherent light to replace the single-photon source, the security and feasibility of BB84 QKD have been widely demonstrated experimentally in fiber~\cite{tanaka2008ultra,yin2008experimental,liu2010decoy,lucamarini2013efficient,frohlich2017long,boaron2018secure}, free space~\cite{Schmitt:2007:Experimental,liao2017satellite} and chip integration~\cite{ma2016silicon,sibson2017chip,Bunandar:2018:Metropolitan} with the help of decoy-state method~\cite{Wang2005Beating,Lo2005Decoy}. To implement the qubit encoding of QKD, one usually has three options: polarization, phase and time-phase. Recent years, the time-phase encoding has received increasing favor in practical system due to two advantages. One is that the reference frame is independent in time basis, which leads a stable and low bit error rate in raw key~\cite{yin2016measurement}. The other is the polarization disturbance immunity in both time and phase bases, which can be deployed in complex field environment while polarization coding systems cannot.

The original BB84 protocol and its security proof~\cite{shor2000simple} directly provide the phase error rate via the total bit error rate of two bases, which is based on the
symmetry of two bases. There are three conditions of this protocol: basis-independent probability selected by Alice, basis-independent probability selected by Bob and the basis-independent detection efficiency, which results in that only half of the raw data can be used to extract key.
In order to satisfy the basis-independent detection efficiency condition, one has to reduce the detection efficiency (including intrinsic loss and efficiency of detector at the receiver) of two bases to be consistent~\cite{liao2018satellite,tanaka2008ultra,yin2008experimental,liu2010decoy,liao2017satellite}, resulting in the lower key rate.

Although QKD can provide fresh secure key in real-time~\cite{liao2018satellite}, the low secret key rate is always the Achilles' heel if one applies the one-time pad encryption. The key rate can be directly doubled at most through the efficient BB84 scheme~\cite{lo2005efficient} without any new technique requirement. The efficient BB84 scheme exploits the biased basis choice and removes the basis-independent detection efficiency condition. To further increase the key rate, combining a biased basis choice with the decoy-state method is proposed ~\cite{wei2013decoy,jiang2016universally,mao2017improved} by an additional basis-independent detection efficiency condition that usually cannot be satisfied and brings security loopholes in the practical system. Recently, a four-intensity decoy-state BB84 QKD~\cite{yu2016reexamination} uses a subtle fact, the expected yields of single-photon component prepared in two bases are the same for a given measurement basis, to remove basis-independent detection efficiency condition and provide a higher key rate. A proof-of-principle experiment~\cite{liu2019experimental} with polarization encoding has shown tens of times key rate improvement in the case of large basis detection efficiency asymmetry. However, the security and feasibility of this protocol are acquired in the finite data with an assumption that Eve is restricted to particular types of attacks. Unfortunately, such assumptions cannot be guaranteed in reality.

In this work, we provide the rigorous finite-key analysis for four-intensity decoy-state BB84 QKD protocol.
The security analysis is based on a combination of entropy uncertainty relation~\cite{tomamichel2012tight,lim2014concise} and a finite-key security bounds~\cite{Yin:2020:tight}. We exploit the autonomous time-phase encoding system to experimentally realize this protocol and continuously distribute secret keys for two months, where there is a 1.8 dB difference between the efficiency of two bases at the receiver. The real-time extracted secret key rate is more than 60 kbps over 50 km single-mode fiber and can be secure against coherent attacks in the universally composable framework~\cite{muller2009composability}.

\section {Coherent security}

We consider a four-intensity decoy-state BB84 QKD protocol, where the basis and intensity chosen with probabilities that are biased. Specifically, the intensities of $\mathsf{Z}$ basis sent by Alice are $\mu$ and $\nu$, the intensity of $\mathsf{X}$ basis is $\omega$, together with the vacuum state without basis information. The bases $\mathsf{Z}$ and $\mathsf{X}$ are selected by Bob with probabilities $q_{z}$ and $q_{x}=1-q_{z}$, respectively. The following is a detailed description of the protocol.

{\it{1.~Preparation.}} Alice exploits the laser to prepare weak pulses with intensities $\mu$ and $\nu$ in $\mathsf{Z}$ basis, $\omega$ in $\mathsf{X}$ basis and the vacuum state given by a random bit $y_{i}$. The probability of selecting intensity $k$ is $p_{k}$ with $k\in\{\mu,\nu,\omega,0\}$. The weak optical pulses go through the insecure quantum channel to Bob.

{\it{2.~Measurement.}}~Bob randomly selects basis $\mathsf{Z}$ and $\mathsf{X}$ with probabilities $q_{z}$ and $q_{x}$ to measure the received pulses, respectively.
Bob records the effective events and corresponding bit $y_i^\prime$.  At least one detector click means an effective event. For multiple detector click, he randomly records a bit value and a basis for passive basis detection.

{\it{3.~Reconciliation.}}~Alice and Bob exploit the authenticated classical channel to announce the effective event, basis and intensity information.
They repeat steps 1 to 3 until $|\cZ_{k}| \geq n^{z}_{k}$ and  $|\cX_{k}| \geq n^{x}_{k}$, where the $\cZ_{k}$ ($n^{z}_{k}$) is set (number) of $k$ intensity prepared by Alice and measured in $\mathsf{Z}$ basis by Bob.

{\it{4.~Parameter estimation.}}~Alice and Bob select a size of $n^{z}_{\mu}+n^{z}_{\nu}$ in $\cZ_{\mu}\cup\cZ_{\nu}$ to get a raw key pair $(\rv{Z}_{\rm A},\rv{Z}_{\rm B})$. They announce the bit value of set $\cX_{\omega}$  and compute the corresponding number of bit error $m_{\omega}^{x}$.
All sets are used to compute the observed number of vacuum events $s_0^{zz}$ and single-photon events $s_1^{zz}$ and the observed phase error rate of single-photon events $\phi_1^{zz}$ in $\rv{Z}_{\rm A}$. If $\phi_1^{zz}\leq \phi_{\rm{tol}}$, they move on to step 5.  Otherwise, they abort the data and start again.

{\it{5.~Postprocessing.}}
Alice and Bob apply an error correction with leaking at most $\lambda_{\rm EC}$ bits of information. They adopt a universal$_{2}$ hash function to perform an error-verification by consuming $\lceil\log_2\frac{1}{\varepsilon_{\rm cor}}\rceil$ bits of information~\cite{wegman1981new}.
At last, they use a universal$_{2}$ hash function to perform privacy amplification on their raw key to get a secret key pair ($\rvS_{\rm A}$,$\rvS_{\rm B}$) with $\ell$ bits.

The four-intensity decoy-state BB84 QKD protocol is $\varepsilon_{\rm cor}$-correct and $\varepsilon_{\rm sec}$-secret in the universally composable framework with~\cite{lim2014concise,Yin:2020:tight}
\begin{equation}\label{eq1}
\begin{split}
\ell = &\underline{s}_0^{zz}+\underline{s}_1^{zz}\left[1-h\left( \overline{\phi}_1^{zz} \right)\right]\\
&-\lambda_{\rm EC}-\log_2\frac{2}{\varepsilon_{\rm cor}}-6\log_2\frac{22}{\varepsilon_{\rm sec}},
\end{split}
\end{equation}
where~$h(x):=-x\log_2x-(1-x)\log_2(1-x)$, $\Pr[\rv{S}_{\rm A}\not=\rv{S}_{\rm B}] \leq \varepsilon_{\rm cor}$ and
$(1-p_{\rm abort})\|\rho_{\rm AE}-U_{\rm A} \otimes \rho_{\rm E}\|_1/2 \leq \varepsilon_{\rm sec}$. Thereinto, $\rho_{\rm AE}$ represents the joint state of $\rv{S}_{\rm A}$ and $\rv{E}$,~$U_{\rm A}$ is the uniform mixture of all possible values of $\rv{S}_{\rm A}$,~and $p_{\rm abort}$ is the probability that the protocol aborts. $\underline{x}$ and $\overline{x}$ denote the lower and upper bounds of observed value $x$.

\begin{figure*}[tbh]
\centering
\includegraphics[width=15cm]{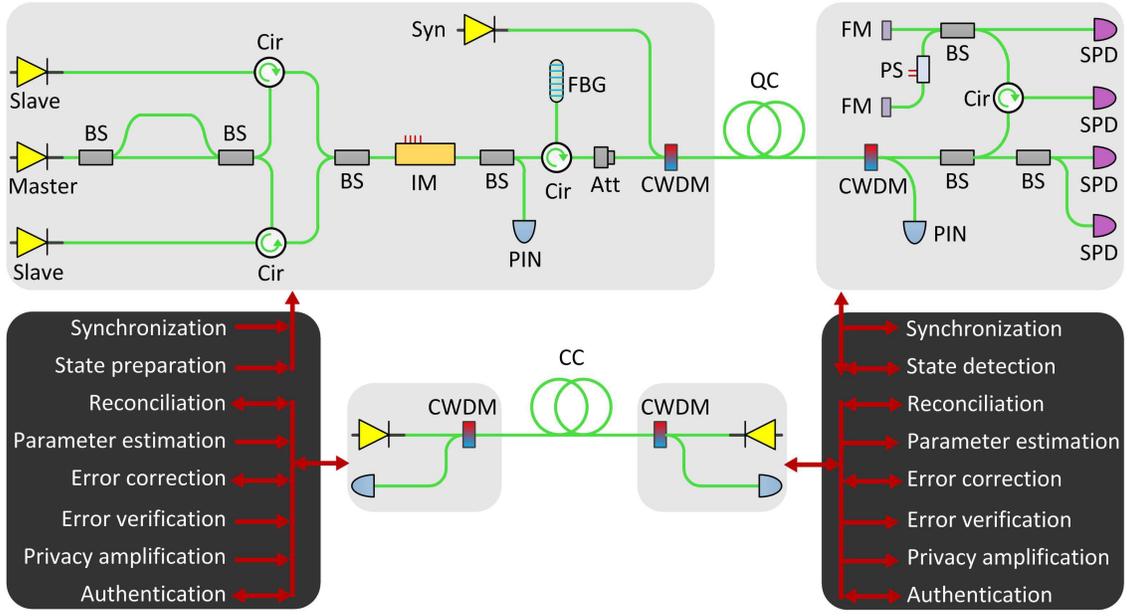}
\caption{Experimental set-up of the decoy-state BB84 QKD system with time-phase encoding. Alice exploits a master laser, two slave lasers and an asymmetric interferometer to prepare optical pulses in $\mathsf{Z}$ and $\mathsf{X}$ basis that are modulated decoy-state using an intensity modulator, before passing through a set of filter, monitor and attenuator to regulate the photon number per pulse. Bob utilizes a biased beam splitter to realize a passive basis detection, following which the pulses either go directly to the time detector or pass through an asymmetric interferometer. A synchronization signal is distributed from Alice to Bob via a wavelength division multiplexed quantum channel. All of the processing is carried out using a FPGA except for the parameter estimation realized in ARM. All classical information is transmitted in a classical channel with an independent optical fiber. BS: beam splitter; Cir: circulator; IM: intensity modulator; FBG: fiber Bragg grating; Att: attenuator; CWDM: coarse wavelength division multiplexer; FM: Faraday mirror; PS: phase shifter; SPD: single-photon detector; QC: quantum channel; CC: classical channel.
} \label{f1}
\end{figure*}

Using the decoy-state method for finite sample sizes~\cite{lim2014concise}, the expected numbers of vacuum event $\underline{s}_0^{zz^*}$ and single-photon event $\underline{s}_1^{zz^{*}}$ in $\rv{Z}_{\rm A}$ can be written as
\begin{equation}
\begin{aligned}\label{eq2}
\underline{s}_0^{zz^*}\geq &( e^{-\mu}p_\mu+ e^{-\nu}p_\nu) \frac{\underline{n}_0^{z^*}}{p_0},\\
\end{aligned}
\end{equation}
and
\begin{equation}
\begin{aligned}\label{eq3}
\underline{s}_1^{zz^*}\geq&\frac{\mu^{2} e^{-\mu}p_\mu+\mu\nu e^{-\nu}p_\nu}{\mu\nu-\nu^2}\\
&\times\left(e^\nu \frac{\underline{n}_\nu ^{z^*}}{p_\nu}-\frac{\nu^2}{\mu^2}e^\mu \frac{\overline{n}_\mu^{z^*}}{p_\mu}-\frac{\mu^2-\nu^2}{\mu^2}\frac{\overline{n}_0^{z^*}}{p_0}\right),
\end{aligned}
\end{equation}
where $x^{*}$ is the corresponding expected value of given observed value $x$, and the upper and lower bounds can be acquired by using the variant of Chernoff bound~\cite{Yin:2020:tight}
\begin{equation}
\begin{aligned}\label{eq4}\nonumber
\overline{x}^{*}&=x+\beta+\sqrt{2\beta x+\beta^{2}},\\
\underline{x}^{*}&=x-\frac{\beta}{2}-\sqrt{2\beta x+\frac{\beta^{2}}{4}}.\\
\end{aligned}
\end{equation}
with $\beta=\ln\frac{22}{\varepsilon_{\rm sec}}$.
The expected number of single-photon event $\underline{s}_1^{xx^{*}}$ in $\cX_{\omega}$ can be given by~\cite{yu2016reexamination}
\begin{equation}
\begin{aligned}\label{eq5}
\underline{s}_1^{xx^*}\geq&\frac{\mu\omega e^{-\omega}p_\omega}{\mu\nu-\nu^2}\left(e^\nu \frac{\underline{n}_\nu ^{x^*}}{p_\nu}-\frac{\nu^2}{\mu^2}e^\mu \frac{\overline{n}_\mu^{x^*}}{p_\mu}-\frac{\mu^2-\nu^2}{\mu^2}\frac{\overline{n}_0^{x^*}}{p_0}\right),\\
\end{aligned}
\end{equation}
where one exploits the fact that the expected yield of single-photon prepared in $\mathsf{X}$ basis is equal to $\mathsf{Z}$ basis given the same measurement basis $\mathsf{X}$.
Besides, the expected number of bit error $\underline{t}_1^{xx^{*}}$ associated with the single-photon event in $\cX_{\omega}$ is~\cite{yu2016reexamination}
\begin{equation}
\begin{aligned}\label{eq6}
\overline{t}_1^{xx}&\leq m_{\omega}^{x}-\underline{t}_{0}^{xx},\\
\end{aligned}
\end{equation}
with
\begin{equation}
\begin{aligned}\label{eq7}
\underline{t}_{0}^{xx^{*}}&=\frac{e^{-\omega}p_{\omega}}{2p_{0}}\underline{n}_{0}^{x^*},
\end{aligned}
\end{equation}
where one utilizes the fact that expected value $m_{0}^{x^*}=n_{0}^{x^*}/2$. For a given expected value, one can use the Chernoff bound to obtain the upper and lower bounds of observed value~\cite{Yin:2020:tight}
\begin{equation}
\begin{aligned}\label{eq8}\nonumber
\overline{x}&=x^{*}+\frac{\beta}{2}+\sqrt{2\beta x^{*}+\frac{\beta^{2}}{4}},\\
\underline{x}&=x^{*}-\sqrt{2\beta x^{*}}.\\
\end{aligned}
\end{equation}
The hypothetically observed phase error rate associated with the single-photon events in $\rv{Z}_{\rm A}$ can be obtained by using the random sampling without replacement~\cite{Yin:2020:tight},
\begin{equation}\label{eq9}
\overline{\phi}_1^{zz}=\frac{\overline{t}_1^{xx}}{\underline{s}_1^{xx}}+\gamma^{U}\left(\underline{s}_1^{zz},\underline{s}_1^{xx},\frac{\overline{t}_1^{xx}}{\underline{s}_1^{xx}},\frac{\varepsilon_{\rm sec}}{22}\right),
\end{equation}
where
\begin{equation}\label{eq10}
\gamma^{U}(n,k,\lambda,\epsilon)=\frac{\frac{(1-2\lambda)AG}{n+k}+
\sqrt{\frac{A^2G^2}{(n+k)^2}+4\lambda(1-\lambda)G}}{2+2\frac{A^2G}{(n+k)^2}},
\end{equation}
with $A=\max\{n,k\}$ and $G=\frac{n+k}{nk}\ln{\frac{n+k}{2\pi nk\lambda(1-\lambda)\epsilon^{2}}}$. The variant of Chernoff bound is used eight times, the Chernoff bound is used four times and the random sampling is used one time. Composing the error terms of finite-sample, we get the factor is 22, including 9 error terms due to the smooth min-entropy estimation~\cite{lim2014concise}.

\section {Experimental realization}

We built a compact and autonomous time-phase encoding QKD system, which continuously distributes secret keys over an optical fiber link using four-intensity decoy-state BB84 protocol secure against coherent attacks. The physical implementation is outlined in Fig.~\ref{f1}. The master laser produces phase-randomized 1.6 ns-wide laser pulses at 1550.12 nm and a repetition rate of 200 MHz. Two pairs of pulses with relative phases 0 and $\pi$ at a 2 ns time delay generated by an asymmetric interferometer are injected into two slave lasers through the optical circulator, respectively. By controlling the trigger electrical signal of two slave lasers, Alice randomly prepares quantum states in $\mathsf{Z}$ (time) and $\mathsf{X}$ (phase) basis using
400 ps-wide laser pulses. The decoy-state scheme is stably implemented by using an intensity modulator, a biased beam splitter with 1:99 and a PIN diode. A fiber Bragg grating with a 50 GHz nominal bandwidth is aligned to remove extra spurious emission and pre-compensate for the pulse broadening in the fiber transmission. The repetition rates of 100 kHz synchronization pulses with 2 ns-wide at 1310 nm are transmitted from Alice to Bob via the quantum channel by using wavelength division multiplexed. The intensities are set as $\mu=0.35$, $\nu=0.15$ and $\omega=0.3$ with the corresponding probabilities $p_{\mu}=0.78$, $p_{\nu}=0.1$ and $p_{\omega}=0.08$.

After the wavelength division demultiplexer, Bob utilizes a biased beam splitter with 3:7 to implement passive choice measurement basis, where the probability of $70\%$ is detected in time ($\mathsf{Z}$) basis and the probability of $30\%$ is detected in phase ($\mathsf{X}$) basis.  The phase variation of the interferometer is compensated by a phase shifter, where the feedback algorithm is completed in advanced RISC machines (ARM). There is a 1.8 dB inherent loss difference between two bases due to the optical element. Four 200 megahertz-gated InGaAs/InP single-photon detectors with effective gate width 450 ps are exploited to detect quantum signals. The efficiency of detector is $20\%$ at a 120 dark count per second. The dead times of detectors in $\mathsf{Z}$ and $\mathsf{X}$ are 3 $\mu$s and 5 $\mu$s, respectively. The detection counts of $\mathsf{Z}$ and $\mathsf{X}$ basis will be affected differently due to the dead time, which also introduces the difference in detection efficiency. The secure and efficient synchronization scheme~\cite{liu2019secure} is used for four gated-mode single-photon detectors calibration.

We continuously run this QKD system for two months over 50.4 km G.652D single-mode fiber with 9.4 dB loss. The real-time extractable secret key rate is always more than 60 kbps in the universally composable framework with $\varepsilon_{\rm sec}=10^{-10}$.
The Winnow algorithm~\cite{buttler2003fast} with 1.42 inefficiency of error correction is used to perform the error correction for a block size of 512 kb. An error verification is carried out after each error correction via the LFSR-based Toeplitz matrix construction~\cite{krawczyk1994lfsr} with 64 bit.  Privacy amplification will be performed by using the concatenation of Toeplitz matrix and the identity matrix ~\cite{hayashi2011exponential} after accumulating data about 4 Mb with error correction ten times,
where the data excludes the amount of information leaked in error correction. We exploit lasers with 1270 and 1290 nm and PIN diodes to realize classical communication with 8B/10B encoding, which will introduce less noise compared with commercial transceiver if quantum and classic channels share one fiber. In order to reduce the noise of synchronization optical pulses, we discard the count of 100 ns before and after each synchronization optical pulse. Authentication, reconciliation, error correction, error verification and privacy amplification are all carried out in a field-programmable gate array (FPGA) on each side, while the parameter estimation is proceeded in ARM.

\section {Conclusion}
In summary, we have experimentally realized the four-intensity decoy-state BB84 QKD with time-phase encoding and exploited tight security bounds for finite-key analysis with composable security against coherent attacks. Our experiment demonstrates that the basis-independent detection efficiency condition has been removed with 1.8 dB difference between time and phase basis of receiver. The stability of our system is very well due to the time-phase encoding. The phase randomness and spectral consistency are guaranteed by the pulsed laser seeding technique~\cite{comandar2016quantum}. Although the secret key rate of 60 kbps is not very high over 50 km fiber. It's enough for some encrypted tasks, for example, voice communication.
Limiting the secret key generation rate is mainly due to the system repetition frequency and the saturation count rate of the single-photon detector, which will be improved in the future.

\section{Acknowledgments}
We gratefully acknowledge support from the National Natural Science Foundation of China under Grant No. 61801420, the Fundamental Research Funds for the Central Universities.

%%%%%%%%%%%%%%%%%%%%%%%%%%%%%%%%%%%%%%%
% choose a style
%\bibliographystyle{ieeetr}
%\bibliographystyle{unsrt}
%\bibliographystyle{naturemag}
\bibliographystyle{apsrev}
%%%%%%%%%%%%%%%%%%%%%%%%%%%%%%%%%%%%%%%

%%%%%%%%%%%%%%%%%%%%%%%%%%%%%%%%%%%%%%%
% choose a .bib file

%\bibliography{Bibli}
%%%%%%%%%%%%%%%%%%%%%%%%%%%%%%%%%%%%%%%

\end{document}